\documentclass[%
 aip,
amsmath,amssymb,
preprint,%
]{revtex4-1}

\usepackage{graphicx}
\usepackage{dcolumn}
\usepackage{bm}
\usepackage[normalem]{ulem}

\usepackage[colorlinks,bookmarksopen,bookmarksnumbered,citecolor=red,urlcolor=red]{hyperref}
\usepackage{amsmath}
\usepackage{graphicx}
\usepackage{natbib}
\usepackage{color}
\usepackage{latexsym}
\usepackage{subfig}
\usepackage{natbib}
\usepackage{upgreek}
\usepackage{mathrsfs}
\usepackage{float}
\usepackage{caption}
\usepackage{soul}
\usepackage{textcomp}
\usepackage{stmaryrd}
\usepackage{amsmath}
\usepackage{lipsum}                     
\usepackage{xargs}                      
\usepackage[pdftex,dvipsnames]{xcolor}  
\usepackage[colorinlistoftodos,prependcaption,textsize=small]{todonotes}

\def\be{\begin{equation}}
\def\ee{\end{equation}}

\def\der#1#2{{\partial #1\over \partial #2}}

\usepackage{verbatim}

\begin{document}

\preprint{AIP/123-QED}

\title[]{Zero absolute vorticity plane Couette flow as an hydrodynamic representation of quantum energy states under perpendicular magnetic field}

\author{E. Heifetz}
\affiliation{Porter school of the Environment and Earth Sciences, Tel Aviv University,
 69978, Israel.}
 \email{eyalh@tauex.tau.ac.il}
 \author{L.R.M. Maas}%
\affiliation{Institute for Marine and Atmospheric research Utrecht, University of Utrecht, 3584 CC Utrecht, NL.}

 \author{J. Mak}%
\affiliation{Dept. of Ocean Science and Center for Ocean Research in Hong Kong and Macau, Hong Kong University of Science and Technology, Clearwater Bay, Hong Kong SAR.}

\date{\today}

\begin{abstract}

Here we extend the Madelung transformation of the Schr\"{o}dinger equation into a fluid-like form to include the influence of an external electromagnetic field on a charged particle. The vorticity of the Madelung fluid is then in the opposite direction to the imposed magnetic field and equal in magnitude to the cyclotron angular frequency. When the particle motion is confined to a plane, perpendicular to an imposed magnetic field, the equivalent flow dynamics is that of zero absolute vorticity obtained in a quasi 2D rotating frame, where the cyclotron frequency plays a role equivalent to that of the Coriolis frequency in a rotating frame. We show how the Landau levels and the extended modes in the integer quantum Hall effect are all mapped into such zero absolute vorticity-like plane Couette flows, where the latter exhibit a geostrophic-like balance between the magnetic force and the gradients of the quantum (Bohm) potential and the electric force.

\end{abstract}

\maketitle

\section{Introduction}

The relation between quantum and fluid mechanics has been established right at the birth of modern quantum mechanics. Less than a year after Erwin Schr\"{o}dinger published his celebrated equation, Erwin Madelung showed (in 1927) that it can be written in a hydrodynamic-like form\cite{Mad27}. This intriguing observation suggested a hydrodynamic approach to quantum mechanics. The latter however gained modest attention in comparison with the dominant Copenhagen interpretation. The Madelung equations (ME) remained also relatively unfamiliar in the fluid dynamics community where the Schr\"{o}dinger equation (SE) is mainly implied as a mathematical formulation to find solutions to the evolution of hydrodynamic waves in different setups\cite{peregrine1983water, maas1997topographic, paldor2007consistent}.

Recently however, an attempt has been made to re-examine and reinterpret fundamental quantum phenomena by analyzing their correspondent hydrodynamic representation. The advantage of this approach is that by mapping quantum phenomena into classical hydrodynamic ones, we recover physical intuition and identify familiar flow patterns that shed a new light on the somewhat counter-intuitive behavior of these phenomena. For instance, from this perspective, quantum tunneling is enabled due to a local balance between the external potential barrier and a pressure gradient force exerted by the Madelung fluid, in a way that the total kinetic (hydrodynamic and internal) energy of the Madelung fluid remains continuous across the potential barrier  \cite{heifetz2020effective}. Another example is the problem of a free falling quantum particle in a gravitational field, that is mapped by the Madelung transform {\color{red}(JM: it is the transform rather than the equations themselves that do the mapping)} into the dynamics of a 1D stably stratified compressible fluid. The quantum energy states are mapped into hydrostatic equilibrium states, where the evaluation of their stability is obtained via the pseudoenergy integral which in that case is the sum of the kinetic and available potential energies \cite{heifetz2020madelung}.

In a recent letter in this journal \cite{HeifetzMass2021Zero} we discussed a mapping between the quantum harmonic oscillator ground state and the zero absolute vorticity plane Couette flow in a rotating frame. 
Here we show that this flow pattern is the generic hydrodynamic mapping of quantum eigen-states with planar channel geometry under an external perpendicular magnetic field. These eigen-states include the Landau levels, as well as the bulk modes of the integer quantum Hall effect. In all these cases the Couette flow results from a geostrophic-like balance between the magnetic force and the gradient of the superposition of the quantum and the electric potentials. 


The paper is organized as follows. 
In Sec.~\ref{sec:2} we present a simple hydrodynamic system, in a rotating frame, admitting a steady solution of zero absolute vorticity plane Couette flow. 
In Sec.~\ref{sec:3} we derive ME in the absence and in the presence of an external magnetic field and draw the analogy with zero absolute vorticity dynamics. Then, in Sec.~\ref{sec:4} we show how the Landau levels and the integer quantum Hall effect extended eigen-states are mapped into the plane Couette flow described in Sec.~\ref{sec:2}. 
We close in Sec.~\ref{sec:5} by discussing the results.   

\section{Zero absolute vorticity plane Couette flow in a rotating barotropic compressible system}
\label{sec:2}

Consider a barotropic, compressible flow in a counterclockwise rotating system, with an angular frequency $\Omega = f/2$, where $f$ is the Coriolis frequency. Viewed from the rotating frame of reference, the flow momentum and continuity equations read: 
\be
{D{\bf u} \over D t} =
-\nabla \left [Q(\rho) + V \right ] - {\pmb f}\times{\bf u}\, ,
\label{Pol_momentum}
\ee
\be
\der{\rho}{t} = -\nabla\cdot(\rho {\bf u})\, .
\label{Pol_cont}
\ee
Here $t$ denotes the time and the nabla operator is defined in the Cartesian coordinates $(x,y,z)$, where $z$ is the vertical coordinate. The velocity field is given by
${\bf u} = (u,v,w)$, the materiald erivative is  $\mathrm{D}/\mathrm{D} t \equiv \left ( \partial/\partial t + {\bf u}\cdot\nabla\right )$, and ${\pmb f}\equiv f {\hat {\bf z}}$, where ${\hat {\bf z}}$ is the vertical unit vector and the axis of rotation. $Q(\rho)$ is the flow enthalpy (so that $-\nabla Q$ is the pressure gradient force), $\rho$ is the density and $V$ is a time independent external potential. 

By defining ${\pmb \omega}_a$ as the absolute flow vorticity (the flow vorticity viewed from a non-rotating frame of rest), which is the sum of the flow vorticity ${\pmb \omega}$ measured in the rotating frame and the vorticity contributed by the rotation of the system:
\be
{\pmb \omega}_a \equiv {\pmb \omega} + {\pmb f}\, , \hspace{0.5cm} {\pmb \omega} = \nabla \times {\bf u}\, ,
\ee
the system \eqref{Pol_momentum}-\eqref{Pol_cont} then satisfies the material line equation for $({\pmb \omega}_a/\rho)$ \citep{kundu}:
\be
{D\over Dt}\left ( {{\pmb \omega}_a\over \rho}\right ) =\left [ \left ( {{\pmb \omega}_a\over \rho}\right )\cdot\nabla\right ]{\bf u}\, .
\label{PV}
\ee
Thus, for a strictly 2D horizontal flow, in a plane perpendicular to the rotation axis, the RHS vanishes and $({\pmb \omega}_a/\rho)$ is materially conserved (which is a direct consequence of the material conservation of circulation in a frame of rest). \eqref{PV} is trivially satisfied for zero absolute vorticity flows. Furthermore, as ${\bf u}\cdot\nabla{\bf u} = {\pmb \omega}\times {\bf u} +\nabla\left (|{\bf u}|^2 /2 \right )$, for zero absolute vorticity \eqref{Pol_momentum} can be written as:
\be
\der{{\bf u}}{t} =
-\nabla \left ({ |{\bf u}|^2 \over 2} + Q + V \right )\, ,
\label{TDBernoulli}
\ee
thus for a steady flow, the time independent Bernoulli equation is satisfied:
\be
{ |{\bf u}|^2 \over 2} + {Q} +{V}  = \mbox{Be} = \textnormal{constant}\, ,
\label{fBernoulli}
\ee
where $\mbox{Be}$ is the Bernoulli potential.

As ${\pmb \omega} = - {\pmb f}$ implies $\left ( \partial v/\partial x - \partial u/\partial y\right ) =  -f$,
the plane Couette flow: 
\be
u(y) = u_0 +f\, y\, ,  \qquad
u_0 = u(y=0)\, ,
\label{fCouette}
\ee
is a simple example of such a zero absolute vorticity flow. When both $\rho$ (hence $Q$) and $V$ are only functions of $y$, \eqref{fCouette} can be a stationary solution of \eqref{Pol_momentum}-\eqref{Pol_cont}, provided that an extended geostrophic balance is maintained between the Coriolis force and the sum of the pressure gradient force and the gradient of the external potential $V$: 
\be
f u  = -\der{}{y}(Q+V) \, .
\label{ExtendGeos}
\ee

In what follows, we show that this simple stationary plane Couette flow is the generic hydrodynamic representation of fundamental quantum eigen-state solutions in a rectangular geometry, on a plane perpendicular to an imposed magnetic field. There, the cyclotron frequency $\omega_c$ plays the role of $f$. To show that, we next derive the hydrodynamic Madelung transformation of the Schr\"{o}dinger equation, first (for completeness) for a neutrally charged quantum particle, and then for a charged particle in the presence of an electromagnetic field.

\section{Madelung transform of the Schr\"{o}dinger equation}
\label{sec:3}


\subsection{ME for a neutrally charged particle}
\label{sec:3A}

The Schr\"{o}dinger equation (SE), for a non relativistic, neutrally charged, spinless quantum particle of mass $m$, in the presence of an external scalar potential $V$ reads\cite{Merz}:
\be 
i\hbar\der{\Psi}{t} = {\hat H} \Psi  = \left ( { {\hat {\bf p}}^2 \over 2m} + m V \right )\Psi \, ,
\qquad
{\hat {\bf p}} = -i\hbar\nabla\, .
\label{Schrodinger}
\ee
Here ${\hat H}$ and ${\hat {\bf p}}$ denote respectively the energy (Hamiltonian) and momentum operators, acting on the particle wavefunction $\Psi({\bf r},t)$, where ${\bf r}$ and $t$ denote respectively the position vector and time. Writing the wavefunction in its polar form:
\be
\Psi({\bf r},t) = \sqrt{\rho}({\bf r},t)e^{i S({\bf r},t)/\hbar}\, ,
\label{Psi}
\ee
then $\rho({\bf r},t)$ is the probability density function (PDF) to find the particle in position ${\bf r}$ at time $t$, and $S$ is the wavefunction phase, scaled by the reduced Planck constant $\hbar$.

As SE is a complex equation, Madelung \cite{Mad27} decomposed  it into its amplitude and phase to obtain two equations. The evolution for the amplitude is:
\be 
\der{\rho}{t} = -\nabla\cdot\left [\rho \nabla \left ({S\over m}\right ) \right ]\, .
\label{cont0}
\ee
Defining then the velocity according to the de Broglie guiding equation \citep{Merz}:
\be
{\bf u} =  \nabla {\tilde S} \, , 
\label{u_potential}
\ee
plays the role of the velocity potential, where tilde denotes hereafter division by the mass particle $m$, ${\tilde S}\equiv S/m$. \eqref{cont0} takes the form of the familiar continuity equation: 
\be 
\der{\rho}{t} = -\nabla\cdot (\rho {\bf u})\, .
\label{cont}
\ee
For the evolution of the phase, Madelung obtained:  
\be
\der{\tilde S}{t} = - \left ({ {\bf u}^2 \over 2} + {Q} + {V} \right )\, ,
\label{HJ}
\ee
where
\be
{Q} = -  {{\tilde {\hbar} }^2\over 2} {\nabla^2\sqrt{\rho} \over \sqrt{\rho}}\, ,
\label{Q_quantum}
\ee
is now the quantum potential (denoted also as the Bohm potential \cite{HeifetzCohen2015}). 
\eqref{HJ} can be regarded as the quantum Hamilton-Jacobi equation where ${\tilde S}$ plays the role of the action and ${Q}$ is the quantum correction (for instance, for the electron mass, ${\tilde \hbar}^2 \sim \mathcal{O}(10^{-8}\, \mathrm{m}^4\, \mathrm{s}^{-2})$). Equivalently, \eqref{HJ} can be regarded as the time dependent Bernoulli equation of the barotropic, inviscid, compressible Madelung fluid where ${Q}$ plays the formal role of its enthalpy \cite{HeifetzCohen2015}. As opposed to classical fluids, this ``enthalpy'' contains spatial derivatives of the density, and is a peculiarity of the Madelung fluid.  

In the absence of quantized vortices resulting from topological defects, the Madelung flow is irrotational, i.e., ${\boldsymbol \omega} = \nabla \times {\bf u} = \nabla \times (\nabla {\tilde S}) = 0$. Taking the gradient of the two sides of \eqref{HJ} (and recalling again that
$({\bf u}\cdot \nabla){\bf u} = {\boldsymbol \omega}\times{\bf u} + \nabla ({|{\bf u}|^2 / 2})$), Madelung obtained the Euler-like momentum equation:
\be
{D{\bf u}\over Dt} = -\nabla\left ({ Q}+ {V}\right) \, .
\label{Euler}
\ee
Hence the Madelung equations \eqref{cont} and \eqref{Euler}, arising as a transform of the Schr\"{o}dinger equation, is formally identified with \eqref{Pol_momentum}-\eqref{Pol_cont} in a non-rotating system 
(${\pmb f} = 0$), where the Madelung fluid density $\rho$ is the PDF of the position of the quantum particle.

For cases where both $\rho$ and $V$ are time independent, the eigenstates of the time-independent SE ${\hat H} {\Psi}=E{\Psi}$ are of the form $S({\bf r},t) = -E\, t +\eta({\bf r})$ (so that ${\bf u}({\bf r}) = \nabla {\tilde \eta}$). These are mapped into stationary anelastic solutions of the Madelung fluid:
\be 
\nabla\cdot (\rho {\bf u}) = 0\, ,
\label{anelastcont}
\ee
\be
{ |{\bf u}|^2 \over 2} + {Q} +{V}  = {\tilde E} = \mbox{Be}\,  .
\label{SEBernoulli}
\ee
Thus, the energy (divided by the particle's mass) of a quantum eigen-state is the Bernoulli potential of the corresponding Madelung fluid. 

\subsection{ME for a charged particle in the presence of an external electromagnetic field}

Consider now SE for a charged particle $q$, in the presence of external magnetic and electric fields, ${\bf B}$ and ${\bf E}$ respectively, where ${\bf B} = \nabla \times {\bf A}$ (${\bf A}$ is the magnetic vector potential), and ${\bf E} =-\nabla \phi - \partial {\bf A}/\partial t$ ($\phi$ is the electric scalar potential) fields. The Hamiltonian operator now reads\cite{Merz}:
\be
{\hat H} = {1\over 2m}({\hat {\bf p}} - q{\bf A})^2 + mV\, , \qquad V= {\tilde q}\phi\, ,
\label{HMag}
\ee
where $q$ is the charge of the particle. For particles with non-zero charge, in order to satisfy the generalized definition of the canonical momentum under the presence of a magnetic field, ${\bf u}$ (the mechanical momentum per mass particle) is redefined as (cf. \eqref{u_potential}):
\be
{\bf u} = \nabla {\tilde S} - {\tilde q}{\bf A} \, . 
\label{v_magnetic}
\ee
With the Coulomb gauge condition $\nabla\cdot {\bf A} = 0$, equations \eqref{cont} and \eqref{HJ} remain unchanged, and a Helmholtz-like decomposition of the Madelung velocity field yields:  
\be
\nabla\cdot {\bf u} = \nabla^2 {\tilde S}\, , \qquad
{\boldsymbol \omega} = \nabla\times {\bf u} = 
-{\boldsymbol \omega}_c\, , 
\label{Helmholtz}
\ee
where $B=|{\bf B}|$, ${\boldsymbol \omega}_c \equiv {\tilde q}\,{\bf B} =({\bf B}/ B)\omega_c\,$, and $\omega_c \equiv {\tilde q}B$ is the cyclotron angular frequency.

Notice that the Madelung fluid divergence field is a variable, but its vorticity field is dictated by the imposed magnetic field. The latter is equal to the cyclotron angular frequency and pointing in a direction opposite to the imposed ${\bf B}$ (where, for an electron with the negative charge $q = -e$, the vorticity is aligned with the magnetic field). The gradient of \eqref{HJ} yields, after rearrangement:
$$
\der{{\bf u}}{t} + \underbrace{\nabla\left ({{\bf u}^2\over 2}\right )}_{({\bf u}\cdot \nabla){\bf u} - {\boldsymbol \omega}\times{\bf u}}  = -\nabla{Q} \underbrace{-\left ( \nabla \phi + \der{{\bf A}}{t} \right )}_{\bf E}{\tilde q}   \, .
$$
Since ${\boldsymbol \omega} = -{\boldsymbol \omega}_c$, we have$ {\boldsymbol \omega}\times {\bf u} = \tilde{q}\,{\bf u}\times{\bf B}$, so adding these terms respectively to both sides, we obtain the Madelung equation in the presence of the Lorentz force {\color{red}(JM: normally textbooks write ${\bf u}\times{\bf B}$ so unless there is a reason for flipping the sign I suggest we stick with convention)}:
\be
{D{\bf u}\over Dt} = -\nabla  {Q} +{\tilde q}\,({\bf E} + {\bf u}\times{\bf B})\, .
\label{Momentum_Schrod0}
\ee
Furthermore, for the case of a time independent magnetic field ($\partial {\bf A} / \partial t = 0$, so that ${\tilde q}\,{\bf E} = -\nabla\, V$), we obtain: 
\be
{D{\bf u}\over Dt} = -\nabla[Q(\rho) +V] - {\pmb \omega}_c\times{\bf u}\, ,
\label{Momentum_Schrod}
\ee
which is equivalent to \eqref{Pol_momentum} with ${\pmb \omega}_c \leftrightarrow {\pmb f}$, where the equivalent condition for zero absolute vorticity (${\pmb \omega} + {\pmb \omega}_c = 0$) is satisfied by \eqref{Helmholtz}.



\section{Plane Couette flow representation of quantum eigen-states}
\label{sec:4}

\subsection{General setup}
\label{sec:4A}

Consider (stationary) eigen-states where the magnetic field is constant and pointing in the $z$ direction, ${\bf B} = B{\hat {\bf z}}$, and the electric field, where it exists, is pointing in the $y$ direction with ${\tilde q}{\bf E}(y) = (-\partial V / \partial y) {\hat {\bf y}}$, and $V = V(y)$. Furthermore, if we take $\rho =\rho(y)$, and consequently $Q=Q(y)$, then for the plane Couette flow ${\bf u} = u(y){\hat {\bf x}}$:
\be
u(y) = u_0 +\omega_c\, y\, , \qquad u_0 = u(y=0)\, ,
\label{Couette}
\ee
we have the extended geostrophic like balance (from \eqref{Momentum_Schrod}):
\be
\omega_c u  = -\der{}{y}\left (Q +V \right ) \, .
\label{Geos}
\ee
These eigen-state solutions differ from each other by their PDF distribution $\rho(y)$ and the energy eigen-value ${\tilde E}$ corresponding to the Bernoulli potential. As quantum energy states are quantized, taking the subscript $n$ to represent those eigen-states where $n=0,1,2...$, the Bernoulli equation \eqref{SEBernoulli} then reads:  
\be
{ (u_0 +\omega_c y)^2 \over 2} + Q[\rho_n(y)] +V(y)  = {\tilde E}_n\, .
\label{YBernoulli}
\ee
Taking the Landau Gauge \citep{tong2016lectures} ${\bf A} = -By{\hat {\bf x}}$,
we obtain that $u_0 = \der{\tilde S}{x}$ from \eqref{v_magnetic}. For each Fourier component in the $x$ direction of the wave function \eqref{Psi} (denoted by subscript $k$), we have $\Psi_{k,n} =\sqrt{\rho_{k,n}(y)}e^{i(kx-E_n t/\hbar)}$ and $u_0 = \hbar k/m$. Thus, the sinusoidal variation in $x$ of the wave function contributes only to a shift in the $y$ direction of the zero velocity line of the Couette flow. Consequently, the hydrodynamic representation of the Fourier components of these eigen-states are exactly the Couette velocity profiles:
\be
u_k(y) = {\tilde \hbar} k +\omega_c\, y\, ,
\label{Landau_Couette}
\ee
or, equivalently: 
\be
u(Y_k) = \omega_c\, Y_k\, , \quad
Y_k = y + L_b^2 k\, , \quad
L_b \equiv \sqrt{{\tilde \hbar}\over \omega_c} = \sqrt{{\hbar}\over q B}\, ,
\label{YLandau_Couette}
\ee
where $L_b$ denotes the magnetic length \citep{tong2016lectures}. 


\subsection{Zero electric field - the Landau levels}
\label{sec:4B}

\begin{figure}
    \centering
    \includegraphics[width=1\textwidth]{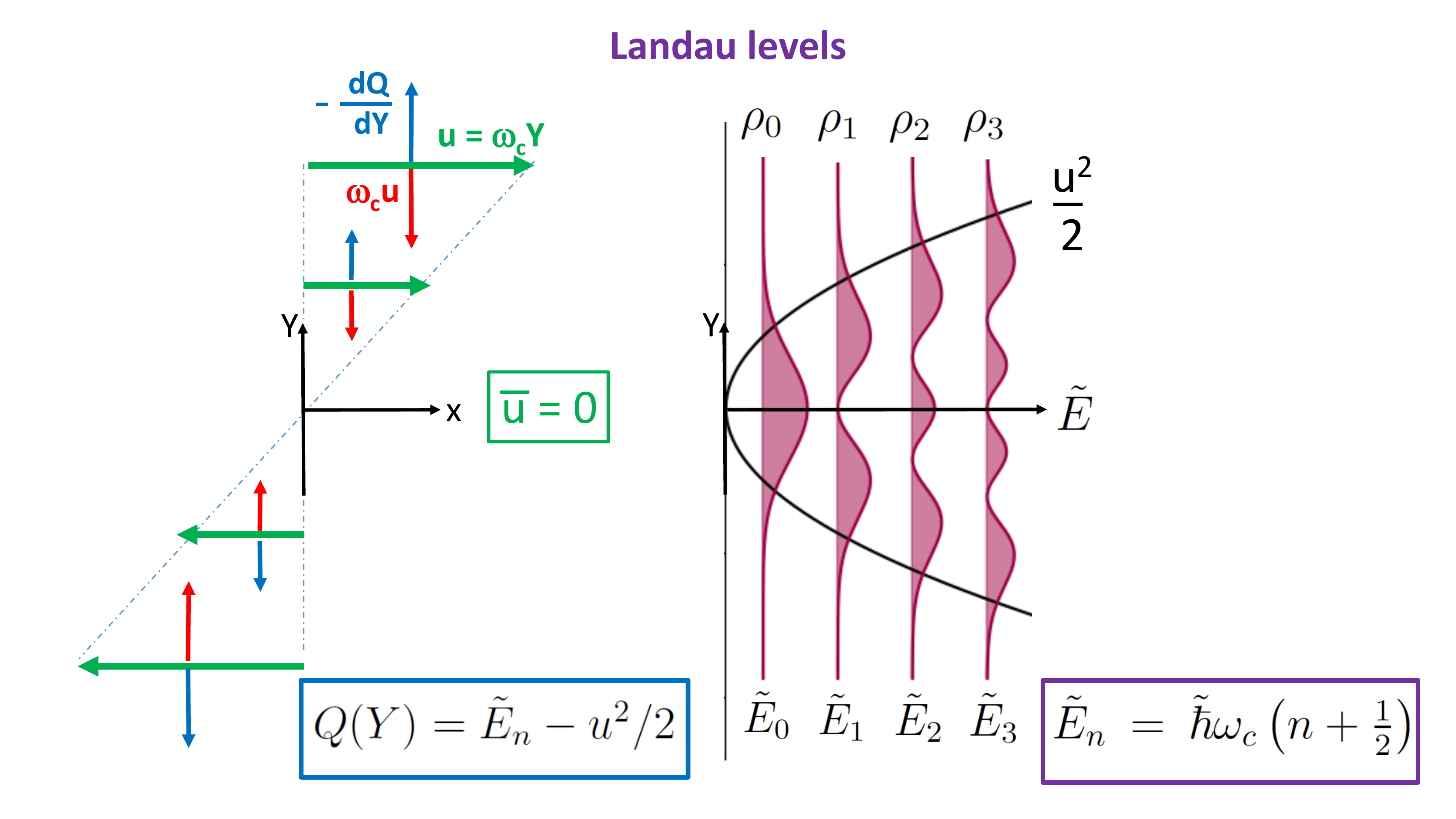}
    \caption{Plane Couette flow representation of the Landau levels. The Couette flow (green arrows) $u = \omega_c Y$ is in the geostrophic-like balance \eqref{PureGeos}, between the magnetic force (red arrows) and the gradient of the quantum potential (blue arrows). The latter satisfies \eqref{Q_quantum} for the Hermite squared polynomials solutions \eqref{Hermite} of the PDFs $\rho_n$ (filled magenta curves), corresponding to the discrete energy levels ${\tilde E}_n$ which are the permitted Bernoulli potential values of the Couette flow. Despite the different structure of $\rho_n(Y)$, $\mathrm{d}Q(Y)/\mathrm{d}Y$ is invariant with $n$, hence satisfying the same geostrophic-like balance for the same Couette flow for all values of $n$. Furthermore, the Landau levels are degnenerate in the sense that for each wavenumber $k$, the structure illustrated in this figure is shifted in the $y$ direction by $L_b^2 k$ (according to \eqref{YLandau_Couette}) without changing the permitted values of the Bernoulli potential. The mean expectation value of the velocity vanishes according to \eqref{UmeanLandau}, as $\rho(Y)$ is symmetric but $u(Y)$ is anti-symmetric.      
    }
    \label{fig:1}
\end{figure}

In the absence of an electric field $V = \textnormal{constant}$, and the geostrophic-like balance \eqref{Geos} and the Bernoulli equation \eqref{SEBernoulli} reduce respectively to (Fig. \ref{fig:1}):
\be
\omega_c u  = -\der{Q}{y} \, ,
\label{PureGeos}
\ee
\be
{ (u_0 +\omega_c y)^2 \over 2} + Q[\rho_n(y)] = {\tilde E}_n\, ,
\label{ZeorEBernoulli}
\ee
where the spatially uniform electric potential is formally absorbed in ${\tilde E}_n$. 


As \eqref{Landau_Couette} and equivalently \eqref{YLandau_Couette} describe the same unbounded plane Couette flows for each $k$, just shifted one from each other in the $y$ direction by $\Delta y =  L_b^2 \Delta k$, it is expected that in the absence of an external potential, such shifts do not change their energy. Therefore all the Fourier components are degenerate in the sense that they possess the same series of possible values of the Bernoulli potential. This can be verified explicitly when substituting \eqref{Q_quantum} in \eqref{ZeorEBernoulli}:
\be
{1 \over 2}\left [ (\omega_c\, Y_k)^2  
 -  {{\tilde \hbar}^2 \over \sqrt{\rho}_n} \frac{\mathrm{d}^2\sqrt{\rho}_n} {\mathrm{d}{Y_k}^2}\right ] = {\tilde E}_n\, ,
\label{HarBernoulli}
\ee
which is identical to the time independent SE for the quantum harmonic oscillator where $\omega_c$ plays the role of the oscillator frequency. Thus the harmonic potential $(\omega_c\, Y_k)^2/ 2$ is mapped into the kinetic energy of the Couette flow \cite{HeifetzMass2021Zero}.

To guarantee localized solutions that vanishes for $|Y_k| \rightarrow \infty$, the harmonic oscillator allows only quantized energy states solutions with $\mbox{Be} = {\tilde E}_n = {\tilde \hbar} \omega_c \left (n + {1\over 2} \right )$ for non-zero positive integer values of $n$, with the corresponding density structure:
\be
\rho_{k,n}(y) = {\rho_{n}({Y_k})}  = {1 \over 2^n n!} 
\sqrt{ \omega_c \over \pi{\tilde \hbar} }\, H_{n}^2 \left (\sqrt{ \omega_c \over {\tilde \hbar}}\, {Y_k}\right )e^{-{ \omega_c {Y_k}^2 / {\tilde \hbar}}},
\label{Hermite}
\ee
where $H_{n}$ are the Hermite polynomials of order $n$, and the multiplicative factors are chosen to normalize the PDF, so that $\int_{-\infty}^{\infty}\rho_n({Y_k}) \mathrm{d}{Y_k} = 1$\citep{Merz}.

The energy levels are the Landau levels. From a hydrodynamic perspective, one may ask for the reason for quantized solutions: why can the Bernoulli potential take on only discrete set of values, $\mbox{Be} = {\tilde E}_n$, for the stationary Couette flow, rather than any general set of continuous values? The reason comes from the peculiar structure of the enthalpy in the Madelung fluid. Enthalpy in the form of \eqref{Q_quantum} can satisfy the parabolic structure ${Q}({Y}) = {\tilde E}_n - {u^2 / 2} = \mbox{Be} - { (\omega_c\, {Y})^2 / 2}\,$ only with specific density structures that correspond to specific discrete values of $\mbox{Be}$. Conversely, in classical hydrodynamic the enthalpy of barotropic fluids is not constrained to be of the form \eqref{Q_quantum}, thus there is no counterpart quantization of $\mbox{Be}$ in the classical setting.

For each energy state $n$, $\rho_n(Y) u(Y) \mathrm{d}Y$ (hereafter dropping the subscript $k$) is the probability to find the charged particle in between $Y-\mathrm{d}Y/2 < Y < Y+\mathrm{d}Y/2$, moving with the velocity $u(Y)$. The mean (expectation) value of the Couette flow ${\overline u}$, for the Landau levels are zero, i.e.:
\be
{\overline u} \equiv \int_{-\infty}^{\infty} \rho_n(y) u(y) \mathrm{d}y = \int_{-\infty}^{\infty} \rho_n(Y)\, \omega_c\, Y \mathrm{d}Y = 0\, ,
\label{UmeanLandau}
\ee
since all of the $\rho_n(Y)$ functions are symmetric but $u(Y)$ is anti-symmetric. This stands in agreement with the geostrophic balance \eqref{PureGeos}, as the mean value of the quantum potential gradient vanishes $\int_{-\infty}^{\infty} \rho_n(Y) (\mathrm{d} Q/ \mathrm{d} Y) \mathrm{d}Y = 0$. This can be verified when substituting ${Q}(Y)$ from \eqref{Q_quantum} and then integrating by parts when recalling that both $\rho_n(Y)$ and $\mathrm{d} \rho_n / \mathrm{d} Y$ vanish at $Y \rightarrow \pm \infty$. It stems from the more general vanishing of the mean value of the quantum potential gradient, $\int \rho\, \nabla Q\, \mathrm{d}\Omega =0$, for PDF satisfying standard boundary conditions ($\rho$ and $\nabla \rho$ vanishing at the boundaries of the volume domain $\Omega$). {\color{red}(JM: minor inconsistency with notation, $V$ is also the potential)}

The result ${\overline u}=0$ is often interpreted as being in agreement with the classical limit of zero drift of the centers of the circular motions of charged particles in the plane perpendicular to a constant imposed magnetic field (positive (negative) charge particles circle clockwise (anti-clockwise) with the cyclotron angular frequency) \citep{tong2016lectures}. This circular motion is equivalent to the inertial circular motion performed by a fluid particle on an ``f-plane'', resulting from a balance between the Coriolis and the centrifugal forces acting on the fluid particle \citep{Holton2004}.

\begin{figure}
    \centering
    \includegraphics[width=1\textwidth]{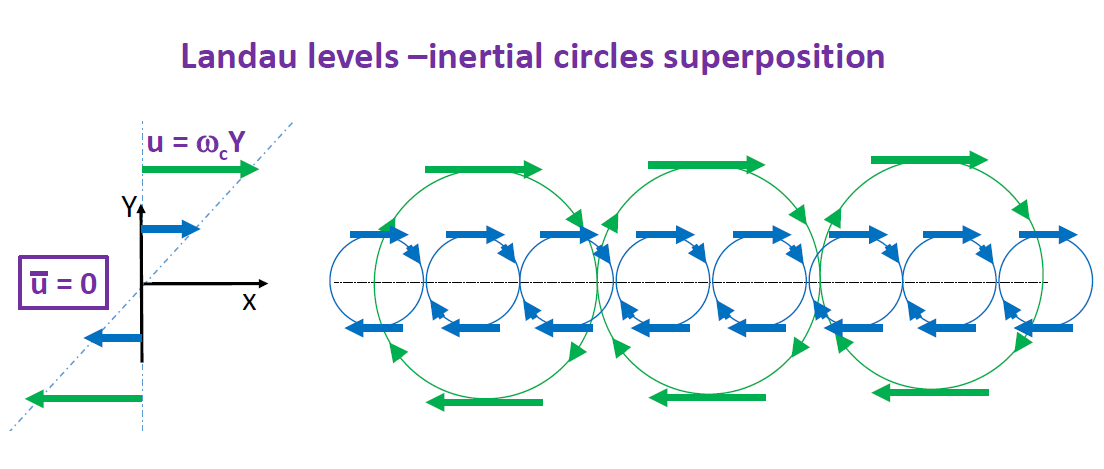}
    \caption{Schematic illustration of how a plane Couette flow can be obtained from a superposition of an infinite number of inertial circles with varying radii, whose centers are located at the Couette zero velocity line, $Y=0$ (see more details in text). 
    }
    \label{fig:2}
\end{figure}

We suggest that we can relate as well a quantum phenomenon in the classical limit to the Couette flow of the Madelung equations {\color{red}(JM: I am guessing this is what the sentence is trying to say)}. Consider a fluid particle at position ${\bf r}(t) = (x(t), y(t))$, circulating clockwise with the cyclotron frequency, around a center ${\bf r}_0 = (x_0, y_0)$, so that the radius of the circle is $|{\bf r} - {\bf r}_0|$. Then the Cartesian components of the fluid particle motion are $u_p = \omega_c(y-y_0)$ and $v_p = -\omega_c(x-x_0)$. Now, consider an infinite number of all possible circles, with all possible radii centered at $y_0 = -L_b^2 k$ ($Y=0$), where $x_0$ varies continuously from $-L_x/2$ to $L_x/2$. Then the velocity field, averaged over $x$, resulting from a superposition of all these circles, is given by $u = (1/L_x)\int_{-L_x/2}^{L_x/2} u_p\, \mathrm{d}x_0 = {\tilde \hbar} k +\omega_c\, y$, which is the Couette flow as in \eqref{Landau_Couette}, and $v = (1/L_x)\int_{-L_x/2}^{L_x/2} v_p\, \mathrm{d}x_0 = -\omega_c\, x$. Thus, at the center of the domain, $x=0$, $v$ vanishes due to the cancellation of positive and negative motion, and in the $y$ direction, every pair of circles whose centers are located at equal distances $|x_0|$, from the left and the right sides of the domain's  center, $x=0$. For an infinite domain ($L_x \rightarrow \infty$), the center can be taken at any point of $x$, so that we can always find corresponding pairs of circular motions whose superposition vanishes $v$ at all $x$. Consequently, the resultant mean superposed flow attributed to all of these possible inertial circles yields the net Couette flow of \eqref{Landau_Couette}; see Fig. \ref{fig:2}.

While in classical fluid dynamics the value of $u_0$ may vary continuously, for the Landau levels $u_0$ is quantized. This leads to the important property of the Landau levels that the areal density of the charged particles is proportional to the magnitude of the perpendicular magnetic field. Each Couette profile $u_k$ of \eqref{Landau_Couette} or \eqref{YLandau_Couette} corresponds to a different charged particle, as each wave number $k= 2\pi/\lambda_x$ is related to the wavelength of the quantum wavefunction of a particle. This number is quantized if we assume periodic boundary conditions at $x=(0,L_x)$, so that  $\lambda_x^j = L_x/j$, for $j=1,2,3...$ Consequently the allowed wavenumbers are $k^j = \left (2\pi /L_x\right ) j$, hence the difference in the $y$ direction between the zero velocity lines of the allowed Couette flows is $\Delta y_0 = 2 \pi (L_b^2 / L_x)$. Thus, the number of allowed charged particles in a slab of width $L_y$ at an energy level is given by $N = L_y / \Delta y_0 = (B / \Phi_0) A$, where $A =  L_x L_y$ is the area of the slab and $\Phi_0 \equiv h / q$ is denoted as the quantum magnetic flux (where $h = 2\pi\hbar$ is the Planck constant). Therefore, the areal charge density in a Landau level $N/A = B/\Phi_0$ (not to be confused with the probability density function, $\rho$, to find a single particle) is proportional to $B$. Consequently, when $\nu$ number of Landau energy levels are filled, the charge areal density $n_{2D} \equiv \nu N/A  = B \nu \Phi_0 = B q \nu /h$.

\subsection{Non zero electric field - the integer quantum Hall effect}
\label{sec:4C}

\begin{figure}
    \centering
    \includegraphics[width=1\textwidth]{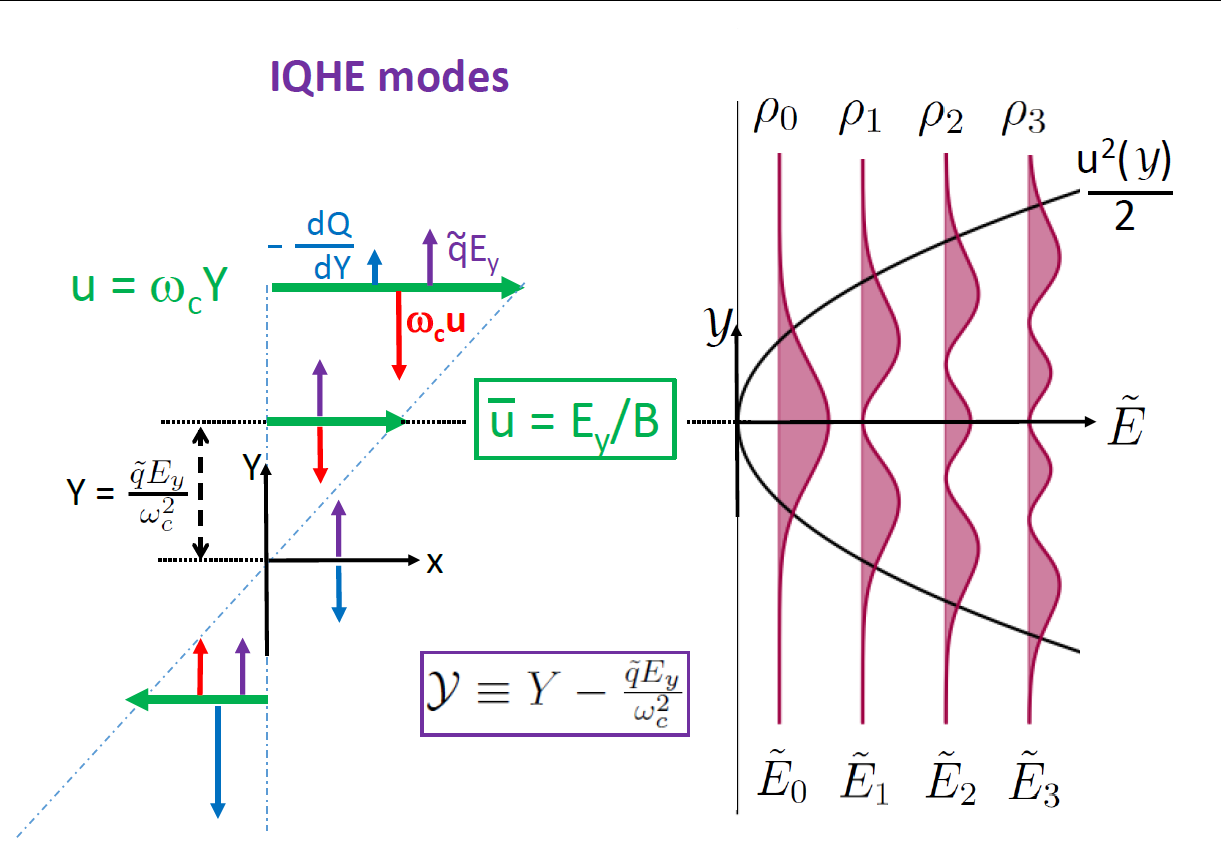}
    \caption{Bulk modes of the integer quantum Hall effect. The setup is the same as in Fig.~\ref{fig:1}, but with an additional constant electric field $E_y$ in the $Y$ direction. The Couette flow is now in a geostrophic-like balance between the gradient of the quantum potential and the magnetic and electric forces (the latter is indicated by the magenta arrows). To maintain the balance, the center of the PDF structures is shifted in the $Y$ direction by ${{\tilde q} E_y\over \omega_c^2}$, in order to vanish the gradient of the quantum potential at the location where the magnetic and the electric forces are in balance. This shift breaks the anti-symmetric relations between $\rho$ and $u$, yielding the non zero expectation value ${\overline u} = E_y/B$ as in \eqref{UmeanPlat}.  
    }
    \label{fig:3}
\end{figure}

When adding a constant transverse electric field ${\bf E} = E_y{\hat {\bf y}}$ ($E_y = -\partial\phi/\partial y$, for $\phi = -y E_y$), the Couette flow solution of \eqref{Landau_Couette} remains unchanged, but the gradient of the quantum potential is now accompanied by the constant electric field to balance the magnetic force, so that \eqref{Geos} reads:
\be
\omega_c u  = -\left (\der{Q}{y} + {\tilde q}E_y \right ) \, .
\label{PlatGeos}
\ee
This corresponds to the extended bulk modes of the integer quantum Hall effect (when $q=-e$). The Bernoulli equation \eqref{SEBernoulli} in the shifted $Y$ coordinate then becomes:
\be
{1 \over 2}\left [ (\omega_c\, Y)^2  
 -  {{\tilde \hbar}^2 \over \sqrt{\rho}} \frac{\mathrm{d}^2\sqrt{\rho}}{\mathrm{d}{Y}^2}\right ]
  -{\tilde q} E_y\, (Y- k L_b^2) = {\tilde E}=\mbox{Be} \, .
\label{PlateBernoulli}
\ee
By making the additional coordinate shift ${\cal Y} \equiv Y  -{\tilde q} E_y / \omega_c^2$, \eqref{PlateBernoulli} can be rewritten, after completing the square and using \eqref{Hermite}, as:
\begin{equation}\begin{aligned}
{1 \over 2}\left [ (\omega_c\, {\cal Y})^2  
 -  {{\tilde \hbar}^2 \over \sqrt{\rho}} \der{^2\sqrt{\rho}}{{\cal Y}^2}\right ] &+ 
 \left ( {E_y\over B}\right ) {\tilde \hbar} k  - {1 \over 2}\left ( {{\tilde q} E_y \over \omega_c}\right )^2 \\
 &= {\tilde \hbar} \omega_c \left (n + {1\over 2} \right )+ 
 \left ( {E_y\over B}\right ) \left [ {\tilde \hbar} k  - {1 \over 2}\left ( {E_y\over B}\right ) \right ]= {\tilde E}_{k,n} =  \mbox{Be}\, .
\label{PlatBernoulli}
\end{aligned}\end{equation}
Hence, the existence of a constant electric field breaks the degeneracy of the Landau levels and makes the Bernoulli constant take different values for each allowed set of $(k,n)$. The reason for the breaking of the degeneracy is demonstrated in Fig.~\ref{fig:3}. While the zero velocity line of the Couette flow is located at $Y_k = 0$, the electric potential zero line is at $Y_k = L_b^2 k$, hence this mismatch depends on $k$. In addition, the density structure $\rho_{k,n}$ of the Hermite polynomials \eqref{Hermite} are centered around ${\cal Y}_k = 0$ so that $Y_k = {\tilde q} E_y / \omega_c^2$, i.e. around $u = \omega_c Y_k = E_y/B$. Therefore, although the Couette velocity profile itself is not altered by the presence of the constant electric field, the shift in the $y$ direction between the density structure and the Couette velocity profile breaks the anti-symmetric structure of $\rho u$ with respect to $Y_k = 0$. Consequently, when multiplying \eqref{PlatGeos} by $\rho$ and integrating over $y$ we obtain that the velocity expectation value is non-zero and satisfying a geostrophic-like balance between the magnetic and the electric forces, where the mean effect of the quantum potential gradient vanishes as before:
\be
\omega_c {\overline u} =  -\der{V}{y} =  \tilde q E_y \hspace{0.25cm} \Longrightarrow \hspace{0.25cm} B{\overline u} = E_y\, .
\label{UmeanPlat}
\ee
The expectation velocity value is equal to the velocity at the PDF center line, regardless of the values of $(k,n)$. It obeys a geostrophic like balance between the magnetic and the electric forces where $E_y/B$ is indeed the classical drift velocity. Hence, the super-positioning of inertial circles, illustrated in Fig. \ref{fig:2}, is applicable as well in the presence of a constant transverse electric field, when viewed from a frame moving with the drift velocity.
 
The expectation value of the charged particle velocity is in agreement as well with the classical limit of the classical Hall effect. Define the Hall voltage difference in the $y$ direction as $V_H \equiv L_y E_y$, then the transverse Ohm's law for the classical Hall effect is $V_H = I_x R_{xy}$, where the electric current in the $x$ direction is $I_x \equiv q\, n_{2D}\, {\overline u}\,L_y$ and the transverse resistance $R_{xy} = B / (q\, n_{2D})$. However, while in the classical Hall effect $R_{xy}$ is proportional to the imposed magnetic field, in the integer quantum Hall effect $n_{2D} = B q \nu /h$, thus $R_{xy} = h / (q^2 \nu)$, is independent of $B$. This results in a series of ``plateaus'' of constant values of $R_{xy}$, when plotted against $B$, in the intervals where the integer number of filled Landau levels remains constant \citep{tong2016lectures}. 

\section{Discussion}
\label{sec:5}

The Madelung momentum equation \eqref{Euler} for a quantum particle is somewhat elusive. It differs from the classical momentum equation by the presence of the gradient of the quantum potential, but its expectation value generally vanishes (when applying $\int \rho\, (...)\, \mathrm{d}V$ on the two sides of the equation). 
Furthermore, for a charged quantum particle in the presence of an electromagnetic force \eqref{Momentum_Schrod}, the velocity field is generally both divergent and rotational, but the latter property is not an independent variable but dictated by the magnetic field. These two subtle issues seem to obscure the simple representation of the energy states of the Landau levels, and the extended modes of the integer quantum Hall effect, as sets of simple plane Couette flows with a constant shear that is equal to the cyclotron frequency. The geostrophic-like balance of this flow together with the Bernoulli energy equation it obeys, provide familiarity and physical intuition from the realm of geophysical fluid dynamics, thus suggesting a different angle of understanding.

This intuition is however partial. There are no direct fluid mechanics counterparts to the quantum effects which result both from the structure of the ``quantum enthalpy'' and the stream-wise quantized boundary conditions of the particle's wavefunction. Nevertheless, it is interesting to compare the magnetic length scale $L_b$ with the Rossby deformation radius length scale $L_d = \sqrt{gH} / f$ (which, roughly speaking, is the length covered in an inertial period $1/f$ by a wave propagating at long wave speed $\sqrt{gH}$, where $f$ is the Coriolis frequency, $g$ is gravity and $H$ is the mean layer thickness). Despite of the enormous scale difference (while $L_d$ varies from $O(10^4-10^6\,\mathrm{m})$ in the ocean and the atmosphere, respectively, $L_b  \approx 2.5 \times 10^{-8}\, \mathrm{m}$ for an electron in a magnetic field of 1 Tesla), if we denote $f = \omega_c \equiv \omega$ and equate  $L_b$ with $L_d$ we obtain $gH = {\tilde \hbar} \omega_c$. Hence, although the quantized energy levels ${\tilde E}_n$ have no direct classical counter-part, the mean potential energy in the geophysical layer plays the role of the energy difference between two adjacent Landau levels, as if jumping from one energy level to the adjacent one requires adding another layer of thickness $H$. Furthermore, in order for $L_d$ and $L_b$ to play an equivalent role, $gH/f$ should correspond to a constant, which means that the mean layer thickness should be proportional to the Coriolis frequency. We cannot think of any reason, or physical constraint to justify it. However, as in shallow water system the fluid density is assumed constant, if such a scenario exists, the mass of a column per unit area, $M/A$, would be proportional to $f$, which corresponds to the quantum case where $n_{2D} \propto B$.

We close the article by noting that while the Madelung equations may sometimes be regarded as a curiosity leading to what might be perceived as superficial links between quantum mechanics and hydrodynamics, there is in fact a deep mathematical connection between the quantum system and hydrodynamics\cite{Khesin-et-al18}. In this case the transform between the Schr\"{o}dinger and Madelung equations is in fact a symplectomorphism of the corresponding phase spaces, and fall under a more general and unified geometric framework with connections to other popular equations in mathematical physics\cite{Khesin-et-al18}. Given the fundamental links, it would be of interest to see whether quantum mechanical features can advance our understanding of fluid dynamics or vice-versa. For the former, we note that the integer quantum Hall effect has been noted have intimate links with topological invariants, and it would be of interest to see how such fundamentally quantum mechanical effects manifest for the analogous fluid system through the bulk-boundary correspondence, some of which have recent received interest in the fluids community\cite{Parker-et-al20a, Delplace-et-al17}. For the latter, we note for example that fluid instabilities have been previously noted to have some formal links with the breakdown of the quantum Hall effect\cite{Eaves01}. Ultimately the goal would be for one field to be able to predict something we do not already know in the other, and to that end further work is required in highlighting links between the two fields through the Madelung formalism, some of which will be our focus in the near future.

\acknowledgements{JM acknowledges financial support from the RGC General Research Fund 16304021 and the Center for Ocean Research in Hong Kong and Macau, a joint research center between the Qingdao National Laboratory for Marine Science and Technology and Hong Kong University of Science and Technology.}

\bibliography{apssamp}

\end{document}